\documentclass[10pt,a4paper]{article}

\usepackage[T1]{fontenc}
\usepackage[utf8]{inputenc}
\usepackage{lmodern}
\usepackage{graphicx,color}
\usepackage{amsmath,amsfonts}%,amssymb,bbm}
\usepackage{tabularx}

\def\A{\mathcal{A}}

\def\H{\mathcal{H}}

\title{ Measures of Entropy and Complexity in altered  states of consciousness}

\author{ D. M. Mateos$^{1}*$, R. Guevara Erra$^{2}$, R. Wennberg$^{3}$, J.L. Perez Velazquez$^{1}$ \\
\footnotesize $^1$~{Neuroscience and Mental Health Programme, Division of Neurology, Hospital for Sick Children.}\\
\footnotesize{ Institute of Medical Science and Department of Paediatrics, University of Toronto, Toronto, Canada.} \\
\footnotesize $^2$~{Laboratoire Psychologie de la Perception, CNRS  and Université Paris Descartes, Sorbonne Paris Cité, Paris, France.} \\
\footnotesize$^3$~{ Krembil Neuroscience Centre, Toronto Western Hospital, University of Toronto, Toronto, Canada. }
\\
\tt\footnotesize * mateosdiego@gmail.com
}

\begin{document}
\maketitle

\begin{abstract}

Quantification of complexity in neurophysiological signals has been studied using different methods, especially those from information or dynamical system theory. These studies revealed the dependence on different states of consciousness, particularly that  wakefulness is characterized by larger complexity of brain signals perhaps due to the necessity of the brain to handle varied sensorimotor information. Thus these frameworks are very useful in attempts at quantifying cognitive states. We set out to analyze different types of signals including scalp and  intracerebral electroencephalography (EEG), and magnetoencephalography (MEG) in subjects during different states of consciousness: awake, sleep stages and epileptic seizures. The signals were analyzed using a statistical (Permutation Entropy) and a deterministic (Permutation Lempel Ziv Complexity) analytical method. The results are presented in a complexity vs entropy graph, showing  that the values of entropy and complexity of the signals tend to be greatest when the subjects are in fully alert states, falling in  states with loss of awareness or consciousness. These results are robust for all three types of recordings. We propose that the investigation of the structure of cognition using the frameworks of complexity will reveal mechanistic aspects of brain dynamics associated not only with altered states of consciousness but also with normal and pathological conditions.

\end{abstract}

%%%%%%%%%%%%%%%%%%%%%%%% INTRODUCTION %%%%%%%%%%%%%%%%%%%%%%%%%%%%%%%%%%%%%%%%%%%%%%%%%%%%%%%%%%%%%%%%%%%%
\section{Introduction}

Multitude of studies focus on the investigation of patterns of correlated activity among brain cell ensembles based on magnitudes of a variety of synchrony indices or similar measures. A prominent common aspect that is emerging from those studies is that of the importance of the variability in the brain coordination dynamics. In general, neurophysiological signals associated with normal cognition demonstrate fluctuating patterns of activity that represent interactions among cell networks distributed in the brain \cite{guevara2016consciousness}. This variability allows for a wide range of configurations of connections among those net-works exchanging information, and thus it supports the flexibility needed to process sensory inputs. Therefore, it has been argued that a certain degree of complexity in brain signals will be associated with healthy cognition, whereas low complexity may be a sign of pathologies \cite{garrett2013moment, velazquez2003dynamical, mateos2014permutation}. We sought to obtain evidence for the correlation between complexity in brain signals and conscious states, using brain electrophysiological recordings in conscious and unconscious states.

There exist a number of statistical measures to analyses electrophysiological recording \cite{hlavavckova2007causality}. In our work we use two well knows measures, one statistical --\textit{Shannon entropy}, a measure of unpredictability of information content in a message \cite{Sha48} and the other deterministic the \textit{Lempel-Ziv complexity} based in the minimum information required to recreate the original signal \cite{LemZiv76}. For both measures, we use use the quantifiers introduced by Bandt and Pompe \cite{BanPom02}, called \textit{permutation vectors}, these are based on the relationship of the neighbour  values  belonging a time series. The Shannon entropy  applying to the permutation vector  is knowing as \textit{permutation entropy} (HPE) \cite{BanPom02}.
%
%and was used for analysis of all types of time series, and particularly has had a major impact on the study of biological signals as the analysis of epilepsy, assessment of anaesthetic efficacy or characterization of  sleep electroencephalograms \cite{ferlazzo2014permutation, olofsen2008permutation, nicolaou2011use}. 
 %
In a similar manner the Lempel-Ziv complexity applied to the permutation vectors is called \textit{permutation Lempel-Ziv complexity} (PLZC) \cite{ZozMat14}. We used these two method to obtain information about the signals dynamics from two differents perspective,  probabilistic (HPE) and  deterministic (PLZC). The permutation entropy and the Lempel-Ziv complexity  have been employed in previous studies analyzing   electrophysiological recording in epilepsy, coma or sleep stages 
\cite{olofsen2008permutation, ferlazzo2014permutation, nicolaou2011use, casali2013theoretically, zhang2001eeg}. Moreover , there is an interesting relation, under certain restrictions,  between the Shannon entropy and the Lempel-Ziv complexity that naturally can extend to the HPE and PLZC \cite{CovTho06, ZozMat14}.

 The result we obtain are shown in a complexity-entropy graph. This kind of representation allows us to visualize better  the results. In a recent  study on chaotic maps and random sequences, it was shown that the complexity-entropy graph  allows the distinction of different dynamics which are impossible to discern using each analysis separately ( \cite{mateos2016PLZCvsHPEsignals}, unpublished results). In our work we  analyze  brain signals recorded using scalp  (EEG),  intracranial electroencephalogram (iEEG) and magnetoencephalogram (MEG), in fully alert states and in two conditions where consciousness is impaired: seizures and sleep. The hypothesis derived from the previous consideration on the variability of activity is that the brain tends towards larger complexity and entropy in wakefulness as compared with the altered states of consciousness.

\section{Methods}

\subsection*{Electrophysiological recordings}

Recordings were analysed from 9 subjects using magnetoencephalography (MEG), scalp electroencephalography (EEG) or intracranial EEG (iEEG). Three epilepsy patients were studied with MEG; one epilepsy patient was studied with iEEG; 3 epilepsy patients were studied with simultaneous iEEG and scalp EEG; and 2 nonepileptic subjects were studied with scalp EEG.

For the study of seizures versus alert states, the three subjects with MEG recordings and the one with iEEG were used. Details of the patients’ epilepsies, seizure types and the recording specifics have been presented in previous studies (MEG patients in \cite{dominguez2005enhanced}, 2005; iEEG patients in \cite{velazquez2011experimental}). For the study of sleep versus alert states, the 3 patients with combined iEEG and scalp EEG have been described previously (patients 1, 3, 4 in \cite{wennberg2010intracranial}); the 2 subjects studied with scalp EEG alone had been investigated because of a suspected history of epilepsy, but both were ultimately diagnosed with syncope, with no evidence of epilepsy found during prolonged EEG monitoring. In brief, the MEG seizure recordings were obtained in one patient with primary generalized absence epilepsy, in one patient with symptomatic generalized epilepsy, and in one patient with frontal lobe epilepsy. The iEEG seizure recordings were obtained from a patient with medically refractory temporal lobe epilepsy as part of the patient’s routine clinical pre-surgical investigation. 

MEG recordings were obtained using a whole head CTF MEG system (Port Coquitlam, BC, Canada) with sensors covering the entire cerebral cortex, whereas iEEG electrodes were positioned in various locations including  the temporal lobe epilepsy patient, the amygdala and hippocampal structures of both temporal lobes. EEG recordings were obtained using an XLTEK EEG system (Oakville, ON, Canada). The details of the acquisitions varied from patient to patient (e.g., acquisition rate varied from 200 to 625 Hz) and were taken into consideration for the data analyses. The duration of the recordings varied as well: for the seizure study, the MEG sample epochs were of 2 minutes duration each, with total recording times of 30-40 minutes; the iEEG patient sample was of 55 minutes duration. The sleep study data segments were each 2-4 minutes in duration, selected from continuous 24-hour recordings.

\subsection*{Data analysis }

The data were analyzed throw the permutation Lempel Ziv complexity (PLZC) and the permutation entropy (PE). Due the relationship existing between these quantities the result were shown in a complexity-entropy graph, to extract information from the signals either deterministic to statistic. In this section we give a breve explanation of both method and the relationship between then.

\subsubsection*{Permutation entropy}

The permutation entropy (HPE) is a measure develop by Bandt and Pompe \cite{BanPom02}, for time series based on comparing neighboring values. The continuous time series is mapped onto a sequence of symbols which describe the relationship between present values and a fixed number of equidistant values at a given past time.

To understand the idea let us consider a real-valued discrete-time series $\{X_{t} \}_{t \ge 0}$ , and let $d \ge 2$ and $\tau \ge 1$ be two integers. They will be called the embedding dimension and the time delay, respectively. From the original time series, we introduce a $d$-dimensional vector $Y^{(d,\tau)}_t$: 
\[
\mathbf{Y}_{t}^{(d,\tau)} \rightarrow  (X_{t-(d-1)\tau} , ..., X_{t-\tau} , X_{t} )  ; t \ge (d - 1)\tau
\]

There are conditions on $d$ and $\tau$ in order that the vector $\mathbf{Y}^{(d,\tau)}_t$ preserves the 
dynamical properties of the full dynamical system\footnote {For EEG signals values of $d = 3,...,7$ have 
been recommended \cite{BanPom02};  For the time lag, it is adequate to use a value of $\tau = 1$  \cite
{bruzzo2008permutation}, For all signals in this work we used the parameter $d=3,..,6$ and $\tau=1$.}. 
The components of the phase space trajectory $\mathbf{Y}^{(d,\tau)}$ are sorted in ascending order. Then, we can define a \textit{permutation vector}, $\Pi^{d,\tau}_t$, with components given by the position of the sorted values of the component of $\mathbf{Y
}^{(d,\tau)}_t$
Each one of these vectors represents a pattern (or motif). There are $d!$ possible patterns. It is possible to calculate the frequencies of occurrence of any of the $d!$ possible permutation vectors. From these frequencies, we can estimate the Shannon entropy associated with the probability distributions of permutation vector. If we denote the probability of occurrence of the  $i$-th pattern  by $P(\Pi^{d,\tau})_i=P_i$ with $i \le d!$ then the (normalized) permutation entropy associated with the time series $\{X_{t} \}$ is (measured in bits):
\begin{equation}
H_{PE}=\frac{-\sum^{d!}\nolimits_{i=1}P_i~log_2 P_i }{log_2 d!}
\end{equation}
The fundamental assumption behind the definition of HPE is that the $d!$ possible permutation vectors
might not have the same probability of occurrence, and thus, this probability might unveil knowledge
about the underlying system.
\subsubsection*{Permutation Lempel-Ziv complexity}

Entropy is a statistical characterization of a random variable and/or sequence. An alternative caracterization of time series is the deterministic notion of complexity of sequences due to Kolomogorof.
In this view, complexity is defined as  the  size of the minimal (deterministic) program (or algorithm) allowing to generate the observed sequence~\cite[Chap.~14]{CovTho06}. Later on,  Lempel and Ziv proposed to define such a  complexity  for the  class  of  ``programs''  based on  recursive  copy-paste operators~\cite{LemZiv76}. 

To  be  more  precise, let  us  consider  a  finite-size sequence  $  S_{1:T}= S_1...S_{T}$  of size  $T$, of  symbols  $S_i$ that  take their  values in  an alphabet $\A$ of finite size $\alpha = |\A|$.  The definition of the Lempel--Ziv complexity lies in the two fundamental concepts of reproduction and production:

\begin{itemize}
\item {\em Reproduction}:  it consists of extending a  sequence $S_{1:T}$ by a
  sequence  $Q_{1:N}$  via recursive  copy-paste  operations,  which leads  to
  $S_{1:_T+N} = S_{1:T} Q_{1:N}$, i.e., where the first letter $Q_1$ is in
  $S_{1:T}$, let us  say $Q_1 = S_i$,  the second one is the  following one in
  the extended sequence of size $T+1$, i.e., $Q_1 = S_{i+1}$ , etc.: $Q_{1:N}$
  is a subsequence  of $S_{1:T+N-1}$. In a sense, all  of the ``information'' of
  the extended sequence $S_{1:T+N}$ is in $S_{1:T}$.
\item {\em  Production}: the extended sequence  $S_{1:T+N}$ is  now such that
  $S_{1:T+N-1}$ can  be reproduced  by $S_{1:T}$, but  the last symbol  of the
  extension can either  follow the recursive copy-paste operation  (thus we face
  to a  reproduction) or  can be ``new''.   Note thus  that a reproduction  is a
  production,  but  the  converse is  false.   Let  us  denote a  production  by
  $S_{1:T} \Rightarrow S_{1:N+T}$.
\end{itemize}

Any sequence can  be viewed as constructed through  a succession of productions, called  a  history  $\H$.  For  instance,  a  history  of $S_{1:T}$  can  be $\H(S_{1:T}): \emptyset \Rightarrow  S_1 \Rightarrow S_{1:2}\Rightarrow \cdots \Rightarrow  S_{1:T}$.  The  number the  productions used  for  the generation $C_{\H(S_{1:T})}$ is in this case equals to the size of the sequence. A given sequence does not have  a unique history and in the spirit  of the Kolmogorov complexity, Lempel and Ziv were interested in  the optimal history, i.e., the minimal number of production  necessary to  generate the sequence.   The size of  the shortest history  is the  so-called Lempel--Ziv  complexity, denoted  as  $C[S_{1:T}] = \min_{\H(S_{1:T})}    C_{\H(S_{1:T})}$~\cite{LemZiv76}. In   a    sense, $C[S_{1:T}]$  describes the  ``minimal''  information needed  to generate  the sequence $S_{1:T}$ by recursive copy-paste operations.

As explained above, the Lempel--Zip complexity ($C_{LZ}$) needed a alphabet of finite size to be used. In continuos time series as EEG or MEG it is necessary to discretized the series before calculating the $C_{LZ}$. Using the same idea that in permutation entropy can be taken the alphabet as the set of permutation vectors $\A= \{\Pi^{(d,\tau)}\}$ and  the alphabet large $\alpha = |d!|$. This is called \textit{permutation Lempel--Ziv complexity} (PLZC)\footnote{
From now we call the permtation Lempel--Ziv complexity as $C_{LZ}$ } \cite{ZozMat14}   

The most interesting thing is although analyzing a sequence from a completely deterministic point of view, it appears that $C_{LZ}[S_{1:T }]$ sometimes also contains the concept of information in a statistical sense. Indeed, it was shown in references \cite{CovTho06, LemZiv76} that for a random stationary and ergodic process, when correctly normalized, the Lempel-Ziv complexity of the sequence tends to the entropy rate of the process; this result were extend to the permutation Lempel-Ziv complexity and the permutation entropy \cite{ZozMat14}; i.e.,
\begin{equation}
\lim_{T \to + \infty} C_{LZ}[S_{1:T}] \frac{\log(T)}{T} \: = \: \lim_{T \to +
\infty} \frac{H_{PE}[S_{1:T}]}{T}
\label{CLZ_HPE:eq}
\end{equation}
where $H_{PE}[S_ {0:T-1} ]$ is the joint permutation entropy of the $T$ symbols, and the righthand side is the permutation entropy rate (entropy per symbol) of the process. Such a property gave rise to the use of the permutation Lempel-Ziv complexity for permutation entropy estimation purposes.
%%%%%%%%%%%%%%%%%%%%%%%%%%%%%%%% RESULT %%%%%%%%%%%%%%%%%%%%%%%%%%%%%%%%%%%%%%%%%%%%%%%%%%%%%%%%%%%%%%%%%%

\section{Results}

The results obtained with recordings acquired during conscious state are compared with those acquired during unconscious states, with include sleep (all stages) and epileptic seizures. We note that while we work at the signal level we made the reasonable assumption that the MEG and scalp EEG sensors record cortical activity underlying those sensors and thus throughout the text we used the term brain signals. On the other hand, the iEEG, obviously, records signals at the source level. For all the signals the permutation vector parameter used were $d = 3,...,6$ and  $\tau=1$.

%%--------------------------------------------------------------------------------------------------------

\subsection{Entropy-complexity analysis from epileptic recordings}

\begin{figure}[H]  
  \begin{center}
    \includegraphics[scale=0.6]{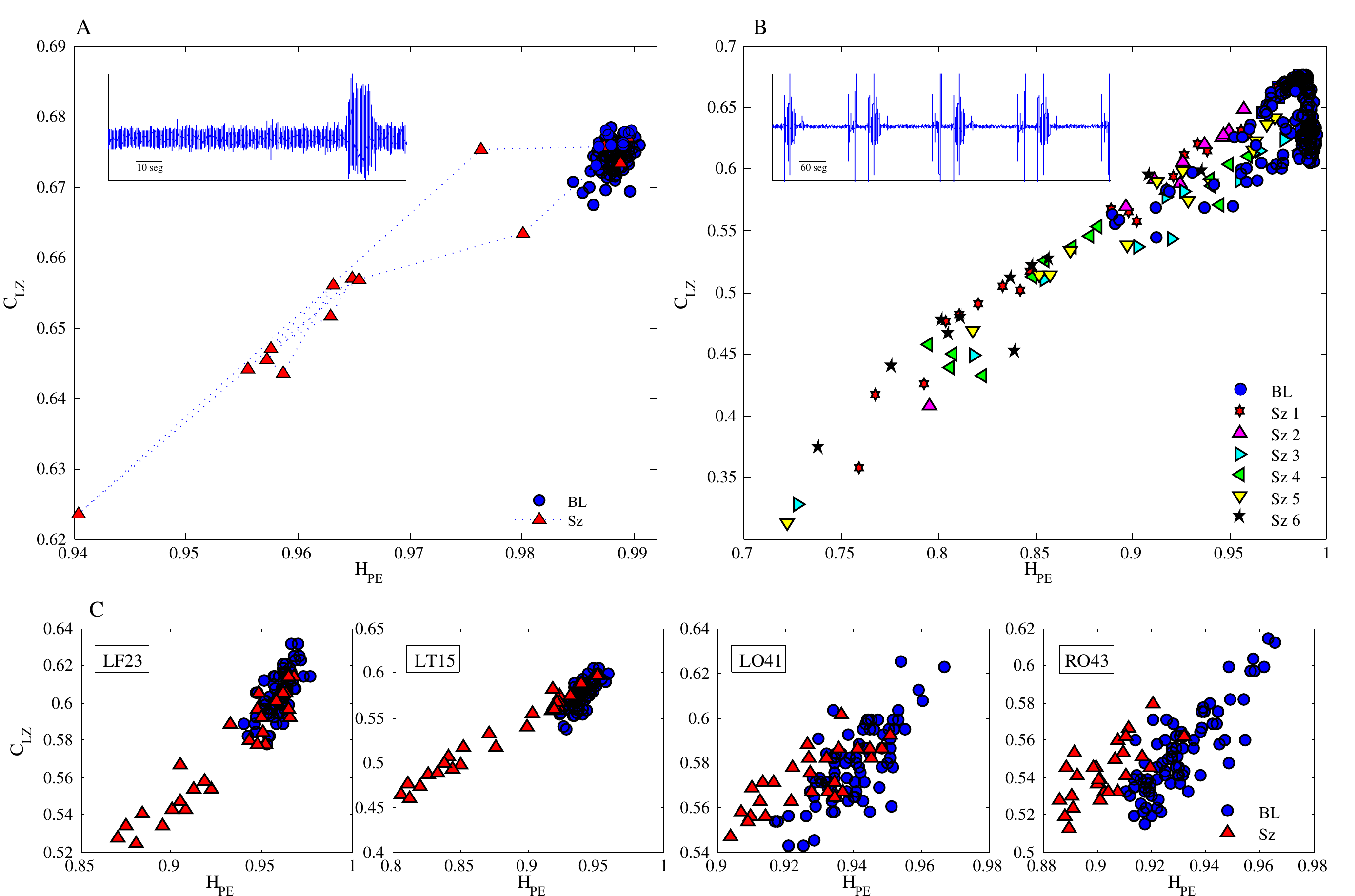}
    
    \caption{  Represent the permutation Lempel Ziv complexity ($C_{LZ}$) vs permutation entropy ($H_{PE}$) (with parameter $d=4$ and $\tau=1$) time tracking values for a MEG signal in epileptic patients during conscious, baseline (BL)  and unconscious, seizure (Sz) states. A) Patient with primary generalized epilepsy, the MEG signal for one channel is plotted in the inset (the high amplitude represent the seizure). We  observe that before the seizure  entropy and the complexity the values remains very high, decreasing in the seizure epoch and return to the original values after the seizure. B) Patient with secondary generalized epilepsy, who had 7 seizure during the recording period, is in the inset. When the patient stay in the inter-ictal state (no seizure period) entropy and complexity values are higher and decreasing in the every attack. C) Patient suffering from frontal lobe epilepsy; 4 channels were analysed separately, left frontal (LF23), left temporal (LT5), left occipital (LO41), right occipital (RO43). For the two recording  areas affecting by the seizure (LF23 and LT5)  entropy and complexity change in the ictal state, but for the areas which are not affected (LO41 and RO43), the $C_{LZ}$ and $H_{PE}$ values are the same that in baseline state. The same result we obtained for the parameter $d=3,4,5,6$ and $\tau=1$.}  
    \label{seizure_figure}
  \end{center}  
\end{figure}

To visualize the  dynamics of entropy and complexity in the time, we use a non overlapping running window ($ \Delta = 625  $) corresponding to $1s$ MEG recording points. For each window the PLZC and HPE were calculated.Fig.~\ref{seizure_figure} shows the complexity (PLZC) and entropy (HPE) values correspond to a MEG recording from a patient suffering primary generalized epilepsy (A), secondary generalized epilepsy (B) and an frontal lobe epilepsy (C). For patients A and B the entropy and complexity values represented were calculated the average over the 143 channels. For  patient C the values in each plot correspond a particular channel. 

One MEG channel corresponding to patient A,  are shown in the inset of Fig.~\ref{seizure_figure}A), where the seizure is visible as a high amplitude in signal. The complexity-entropy graph depict clearly the dynamics of the ictal event. During conscious states (baseline) -- when patients remain conscious since they are not having generalized seizures-- the PLZC and HPE tent to  maximum values, but as the patients experience  seizures both values decrease widely, returning to the original baseline values after the event.

Similar result can be seen in patient B who had 7 seizures, the seizures  are visible in  the inset of Fig.~\ref{seizure_figure}B. We can see in the graph that the baseline and the interictal activity -- the recording between to seizures -- reach always the highest values in entropy and complexity,  declining to values well below in the ictal state (seizure). This result is repeated for each of the seizures.

In Fig.~\ref{seizure_figure}C we show the analysis for 4 different MEG channels corresponding to: left frontal (LF23), left temporal (LT5), left occipital (LO41) and right occipital (RO43). The first two belong to the region where the seizure spread. For all channels the values of HPE and PLZC are higher in baseline, however  the entropy and complexity decay in the most affected areas (LF23, LT5), while for the other areas (LO41, RO43) the complexity doesn't change, there being  a small decrease in entropy. Similar result were found in the signals of the other epileptic patients, recorded with scalp EEG and iEEG.

A possible explanation for this decreas in complexity and entropy during seizures, is that there is higher synchrony during ictal periods (seizures), therefore this causes  the recording signals become more stereotyped, the number of permutation vectors used to quantized the signals are smaller and more regular giving a lower entropy and complexity. This will be further commented in the discussion.

%%--------------------------------------------------------------------------------------------------------

%%% Sleep 
%%----------------------------------------------------------------------------------------------------------

\subsection{Entropy-complexity analysis during sleep stages}

\begin{figure}[H]  
  \begin{center}
    \includegraphics[scale=0.6]{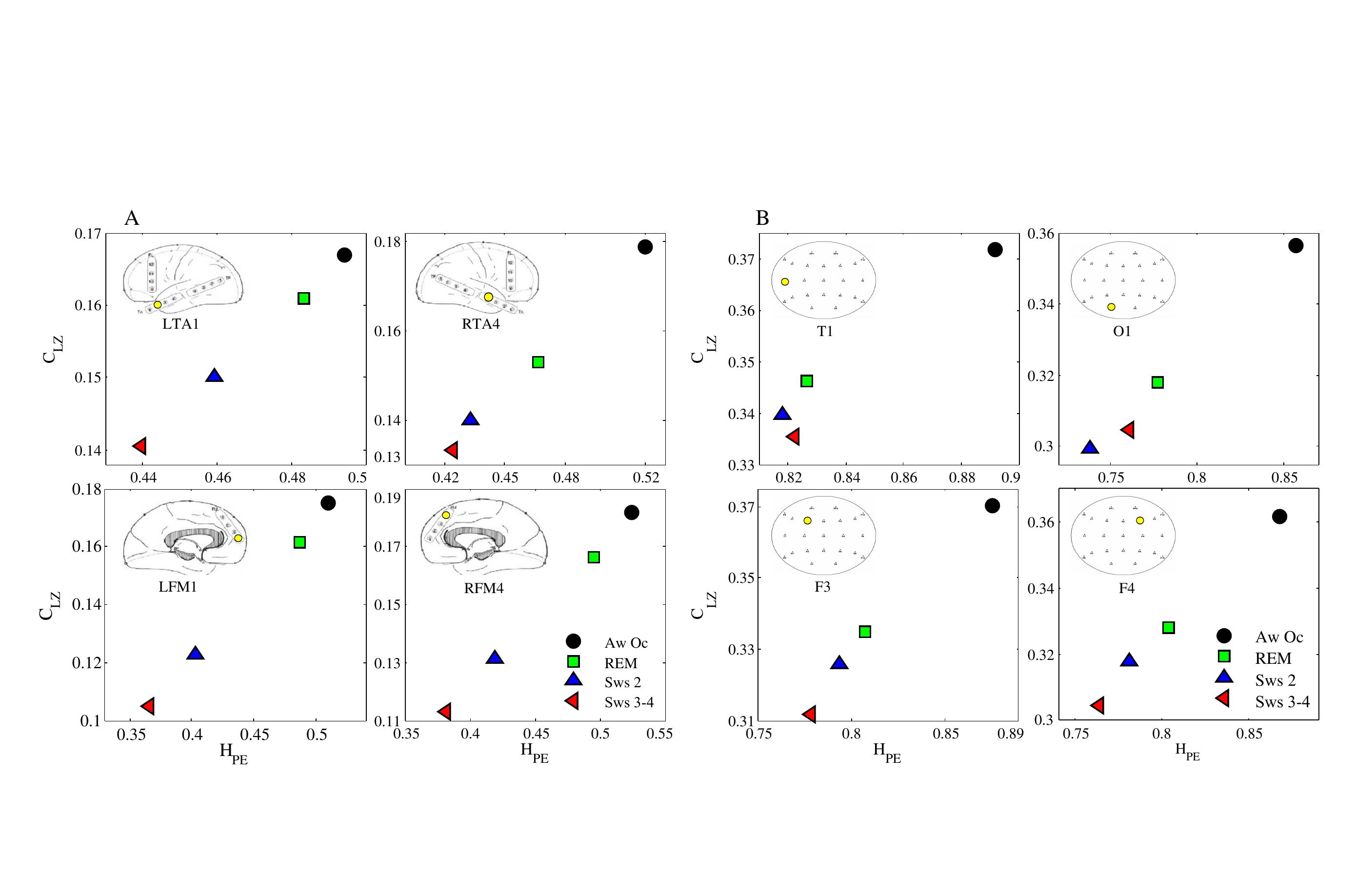}
    
    \caption{ A) Each windows the channel recording from one iEEG channels analyzing by the permutation Lempel-Ziv complexity ($C_ {LZ}$) vs permutation entropy ($H_{PE}$) graph (with parameter $d=4$ and $\tau=1$), for a patient recording during sleep.  Data samples were of 2-4 minutes duration during wakefulness with eyes open ('Aw Oe') , and sleep stages slow-wave 2 (‘Sws2’), slow-wave 3-4 (‘Sws3-4’) and rapid eye movement ('REM'). The electrode localization are: left frontal media (LFM1), rigth frontal media (RFM4), left temporal anterior (LTA1), right temporal anterior (RTA4),  the yellow circle show the position of the channel in the brain. When the patient go in deeper sleep states,  both PLZC and HPE decreases across all channels. B) The same analysis as in A applied to another subject (scalp EEG channel recording), as in the previous one, awake state has the higher entropy and complexity and the values decrease  for deeper state of sleep. For the REM stage the values are remains between the sleep stages and the wakefulness stage. The same result we obtained for all patient analyzed with the parameter $d=3,..,6$ and $\tau=1$.}
  \label{Sleep_figure}  
  \end{center}
  
\end{figure}

Te recording in these cases were of 2-4 minutes duration during wakefulness with eyes opened ('Awake') or closed, and in sleep stages slow-wave 2 (Sws2), slow-wave 3-4 (Sws3-4) and rapid eye movement ('REM'). Fig.~\ref{Sleep_figure}A shows  entropy and complexity values applying to 4 whole recording  (iEEG channels): left frontal media (LFM1), right frontal media (RFM4), left temporal anterior (LTA1), right temporal anterior (RTA4). The various stages of sleep are remarkably differentiated in the graph. Note how during wakefulness entropy and the complexity is in the higher region of the graph, whereas for the slow wave stages, the values stay in the lower region. The deepest sleep stage, slow wave 3-4 (sws 3-4), has consistently the lowest entropy and complexity. Interestingly,  entropy during REM sleep is very close, in most cases, to the normal, alert state. This result may not be as surprising as it appeare, if we consider the mental activity during REM episodes that are normally associated with dreams. The results are in agreement with those reported in \cite{nicolaou2011use, casali2013theoretically}.  

The results for 4 scalp EEG channels  are shown in Fig.~\ref{Sleep_figure}B, where the same result was obtained: higher complexity and entropy for awake state and lower for deep sleep state. In this case, during REM, the values remains between slow-wave period and wakefulness. For the other 2 subjects  analyzed we obtained similar results. This example demonstrates that the same qualitative result is obtained with different recording techniques. The similarity of the result indicate that these type of analysis is not influenced by the recording methodology.

%%------------------------------FIGURE 2------------------------------------------------------------------

%%%%%%%%%%%%%%%%%%%%%%%%%%%%%%%% CONCLUTION %%%%%%%%%%%%%%%%%%%%%%%%%%%%%%%%%%%%%%%%%%%%%%%%%%%%%%%%%%%%%%

\section{Discusion}
Our results indicate a pronounced loss of entropy and complexity in brain signals during unconscious states or in states that do not represent full alertness (eyes closed). This is consistent with what the signals represent: the coordinated collective activity of cell ensembles, which, in alert states, are responsible for optimal sensory processing. This optimality requires certain variability in the interactions among those cell networks, which will be conceivably represented in greater complexity. 

Previous work has indicated less variability in the coordinated activity patterns in altered states of consciousness, mainly derived from the analysis of synchronization in patients in coma \cite{nenadovic2008fluctuations, nenadovic2014phase}, or during seizures \cite{dominguez2005enhanced, velazquez2007complex}. A common feature of several theories of consciousness is the notion of a broad distribution of cellular interactions in the brain that results in conscious awareness (reviewed in \cite{klink2015theories}). This requirement implies that a certain, high degree of variability in the formation and dissolution of functional cell ensembles should take place \cite{flohr1995sensations}, and this variability will be reflected in higher complexity of the brain signals during alert states. Moreover in   several computational studies have revealed as well  the lower complexity associated with epilepsy and abnormal cognitive states, like schizophrenia \cite{steinke2011brain}

In fully alert states, brain recordings exhibit higher frequencies of relatively low amplitude, and are less regular than  during other states where alertness is perturbed, including closing the eyes (when a prominent periodic alpha rhythm appears in parieto-occipital areas, for instance).  Brain cell ensembles that need to integrate and segregate sensorimotor transformations while  they receive  rich sensory-motor inputs \cite{tononi2004information}; it is then conceivable that these characteristics will be reflected in the high entropy and complexity values we observe. As consciousness is gradually lost, during sleep, the values of entropy and complexity decrease because brain networks do not need the richness in states needed to process the sensorium. The lack of arrival of multiple sensory inputs during unconscious states decreases the need for neurons to display many different firing frequencies, since there is not much integration/segregation being done at those stages and there is not much sensory load. One consequence of this change in firing patterns during unconscious states, particularly in sleep (for a comprehensive review of the neurophysiological mechanisms leading to slow-wave sleep and other thalamocortical phenomena see \cite{destexhe2001thalamocortical}) is that the high frequencies (gamma range) become less prominent and there is higher synchrony at lower frequencies. As well, the amplitude of the slow waves is now high since there are more synchronized cells. Thus, all these events result in the recording becoming more regular and exhibiting the typical slow wave frequencies, and therefore our complexity measures decrease as compared to alert states. These results are consistent with measures obtained from analysis of sleep EEG using permutation entropy \cite{nicolaou2011use} and other nonlinear measures, such as approximate entropy, correlation dimension, recurrence plots and Hurst exponent, amongst others \cite{roschke1992nonlinear, acharya2005non, burioka2005approximate}.

In the case of the epileptic recordings we have observed that the complexity and entropy values are larger in the interictal stage (between seizures) and decline sharply in the ictal stage (seizures). This may be due to the fact that seizures are characterized by excessive synchronous neuronal activity, which generates predominance of large amplitude waveforms, the frequencies depending on the seizure type; e.g., the frequency is low in absence seizures (3-4 Hz), but vary substantially in temporal lobe seizures.  However, the frequencies remain relatively constant for certain time periods (originating a distribution of periodic epochs, or laminar phases), that have been used in the characterization of dynamical regimes in epileptiform activity \cite{velazquez1999type}, and therefore the complexity and  entropy tend to decrease. During the sleep stages we also found decreased entropy and complexity as compared with alert states, a reflection of the aforementioned emergence of highly synchronous cell activity during slow wave sleep. On the other hand, we found that complexity during REM sleep is similar to that of the awake state. This is conceivable since REM episodes are normally associated with dreaming, and there is certain cognitive activity going on in dreams, when there is partial awareness. Previous work has shown decreases in HPE and  LZC in patients under anesthesia effects \cite{zhang2001eeg, olofsen2008permutation, li2010multiscale}, thus the decreased complexity of brain signals in unconscious states may be a common phenomenon.

Hence in the final analysis what we measure, at the macro(meso)scopic level (through the recording of collective cel activity in EEG or MEG), is a reflection of that the brain handles more information during wakefulness. A larger code is required to manipulate more information. The complexity/entropy of the signals used in this work have been quantified through the Bandt and Pompe method \cite{BanPom02}, which focuses on the relative values of neighbouring data points in a time series. Every embedding vector (or motif $\Pi^{d,\tau}_i$) gives an idea of how  the waveform is, in a small section, of the original signal. As the original signal carries more variable information, the waveform tend to be more fluctuating, and the number of distinct motifs required to map it increases. Because of that the probability distribution of motifs  $P(\Pi^{d,\tau})$   tends to be uniform, and this caused entropy increase. Besides, due to of the waveform fluctuation, the PLZC increases too, since  much more information is required  to reconstruct  the signal. In contrast, for monotonal repetitive signals which have little new information, just a limited number of motifs are required, e.g. for a sinusoidal signal the PLZC and HPE  tend to be zero.

We note that our present resutls are complementary to those recently obtained using measures of coordinated activity, namely the number of configurations of connections derived from an index of phase synchronization \cite{guevara2016consciousness}; we should consider that the present analysis, done on the raw signals, represent too correlated activity as each local field potential (in case of iEEG) or signals recorded in scalp EEG or MEG represent the collective activity in cell ensembles, thus these signals  are themselves a measure of coordinated cell activity, and therefore it is not surprising we obtain similar observations

It can be concluded that in the awake state, when information has to be  handled is larger, the complexity and
entropy of the signals recorded from the brain tend to be higher than in absence of consciousness, a result that
stems from the distinct waveforms recorded in these mental states.

% ==================== Bibliography ==================== %

\bibliography{Biblio_HvsCLZ_consciousness2}
\bibliographystyle{unsrt}
\end{document}